\documentclass[12pt,preprint]{aastex}
\usepackage{graphicx}
\usepackage{amssymb}
\usepackage{amsmath}
\newcommand{\psr}{PSR J1537+1155}

\shorttitle{Interaction in the DNSB J1537+1155}
\shortauthors{Durant et al.}

\begin{document}

\title{
Orbital variation of the X-ray emission from the double neutron star binary J1537+1155}

\author{Martin Durant\altaffilmark{1}, Oleg Kargaltsev\altaffilmark{1},
Igor Volkov\altaffilmark{2}, and George G.\ Pavlov\altaffilmark{2,3}}

\altaffiltext{1}{Department of Astronomy, University of Florida, Gainesville, FL 32611--2055, USA; martin.durant@astro.ufl.edu, oyk100@astro.ufl.edu}
\altaffiltext{2}{Department of Astronomy and Astrophysics, Pennsylvania State University,
University Park, PA 16802, USA; pavlov@astro.psu.edu}
\altaffiltext{3}{St.-Petersburg State Polytechnical University, Polytekhnicheskaya ul.\ 29, 195251, Russia}


\begin{abstract}
We  observed the double neutron star binary (DNSB) containing \psr\ 
(also known as B1534+12) with the {\sl Chandra} X-ray Observatory. This is one of the two DNSBs detected in X-rays  and the only one where a hint of  variability  with orbital phase was found (in the previous {\sl Chandra} observation). Our follow-up observation 
 supports the  earlier result: the distribution of photon arrival times with orbital phase again shows a deficit around apastron. The  significance of the 
 deficit in the combined dataset exceeds 99\%. Such an orbital light-curve suggests that the X-ray emission is seen only when neutron star B passes through the equatorial pulsar wind of neutron star A.  We 
  describe statistical tests  we used to determine the significance  of the deficit, and conclusions that can be drawn from its existence, such as interaction of the pulsar wind with the neutron star companion. We also provide better constrained spectral model parameters  obtained from the joint spectral fits to the data from both observations. A power-law successfully fits the data, with best-fit photon index $\Gamma=3.1\pm0.4$ and unabsorbed flux $f = (3.2\pm0.8)\times10^{-15}$\,erg\,s$^{-1}$cm$^{-2}$ (0.3--8\,keV range).
\end{abstract}
\maketitle

\section{Introduction}
Double neutron star binaries (DNSBs) are rare systems  comprised of two neutron stars (NSs) in a tight orbit.
In addition to the gravitational interaction, in which general
relativistic effects often play a significant role, the two NSs
 may also interact in other ways (e.g., via electromagnetic torque or pulsar wind). 
  The manifestations of such interaction(s)  can be found via precise radio timing, 
  provided that one of the NSs is seen as a radio pulsar. In the exceptional case of DNSB J0737$-$3039 (hereafter J0737; see \citealt{2008ARA&A..46..541K} for recent review), both the NSs are pulsars,  
which has enabled a suite of high-precision measurements of post-Keplerian parameters \citep{2006Sci...314...97K}, determination of the component masses \citep{2004Sci...303.1153L}, and analysis of the interaction between the wind of pulsar A and the  magnetosphere of pulsar B \citep{2004MNRAS.353.1095L}.
The evolutionary processes in which DNSBs may form have been discussed 
by, e.g., \citet{2005esns.conf..191V}.

J0737 had been the only DNSB detected in X-rays \citep{2004ApJ...605L..41M}, until \citet[hereafter KPG06]{2006ApJ...646.1139K}  discovered  X-ray emission from DNSB J1537+1155 (hereafter J1537). 
KPG06
 observed J1537 with the 
Advanced CCD Imaging Spectrometer (ACIS) on board {\sl Chandra}. In the 36\,ks exposure, they detected 16 photons, 
 whose energies suggested a steep spectrum (photon index  $\Gamma\approx3$). Intriguingly, KPG06 found that the arrival times did not appear to be evenly distributed in orbital phase, showing a deficit of photons near apastron. However, the statistical significance of the deficit in the first observation  was marginal (in fact, as we 
 show below, it was overestimated by KPG06).
    Here we present analysis of the second {\sl Chandra} ACIS
observation (with a re-analysis of the first one), 
which provides additional support for the dependence of its emission on orbital phase.

In Section 2, we describe the data analysis, with details about the timing (\S2.1) and the spectrum (\S2.2). We present conclusions and a brief summary in Section 3.

\section{Observation and Analysis}
In both observations of J1537, the ACIS was operated in the Very Faint (VF) 
mode.
For each of the datasets, we applied VF-mode background filtering \citep{2003SPIE.4851...28G},  removed positional randomization applied by the pipeline,  and  applied the sub-pixel resolution tool to split-pixel 
events  \citep{2001ApJ...554..496T,2001ASPC..251..576M}.  

The first observation (ObsID 5591) was carried out on 2005 April 10, 
with a total exposure of 36080\,s (after filtering for good time intervals).
The source was imaged on-axis on the most sensitive S3 chip, operated in
 Timed Exposure mode with a frame time of 3.24104\,s (full-chip readout, giving 457\,s dead-time and 35623\,s 
net live time).  We reprocessed 
 this data set 
and generated a new
 Level 2 event file\footnote{See {\tt http://cxc.harvard.edu/ciao/threads/createL2/index.html}}, 
in order 
to  ensure consistency with a more recent  pipeline version (DS  8.2) used to process the second data set.
 As a result of slightly different processing, the extraction aperture of $r=1\farcs23$ 
  (used in KPG06) now includes 17 photons (one extra photon compared to KPG06)
  in the energy range of 0.3--10 keV.
For spectral analysis, photons were extracted using the CIAO task {\tt psextract} (CIAO version 4.2, CALDB version 4.2.0). Counting the number of background counts in an annulus  
    ($4\farcs92<r< 44\farcs28$) centered on J1537, we expect to find   (on average) 
  0.37  background photons within the source extraction aperture.

\begin{figure}
\begin{center}
\includegraphics[width=\hsize]{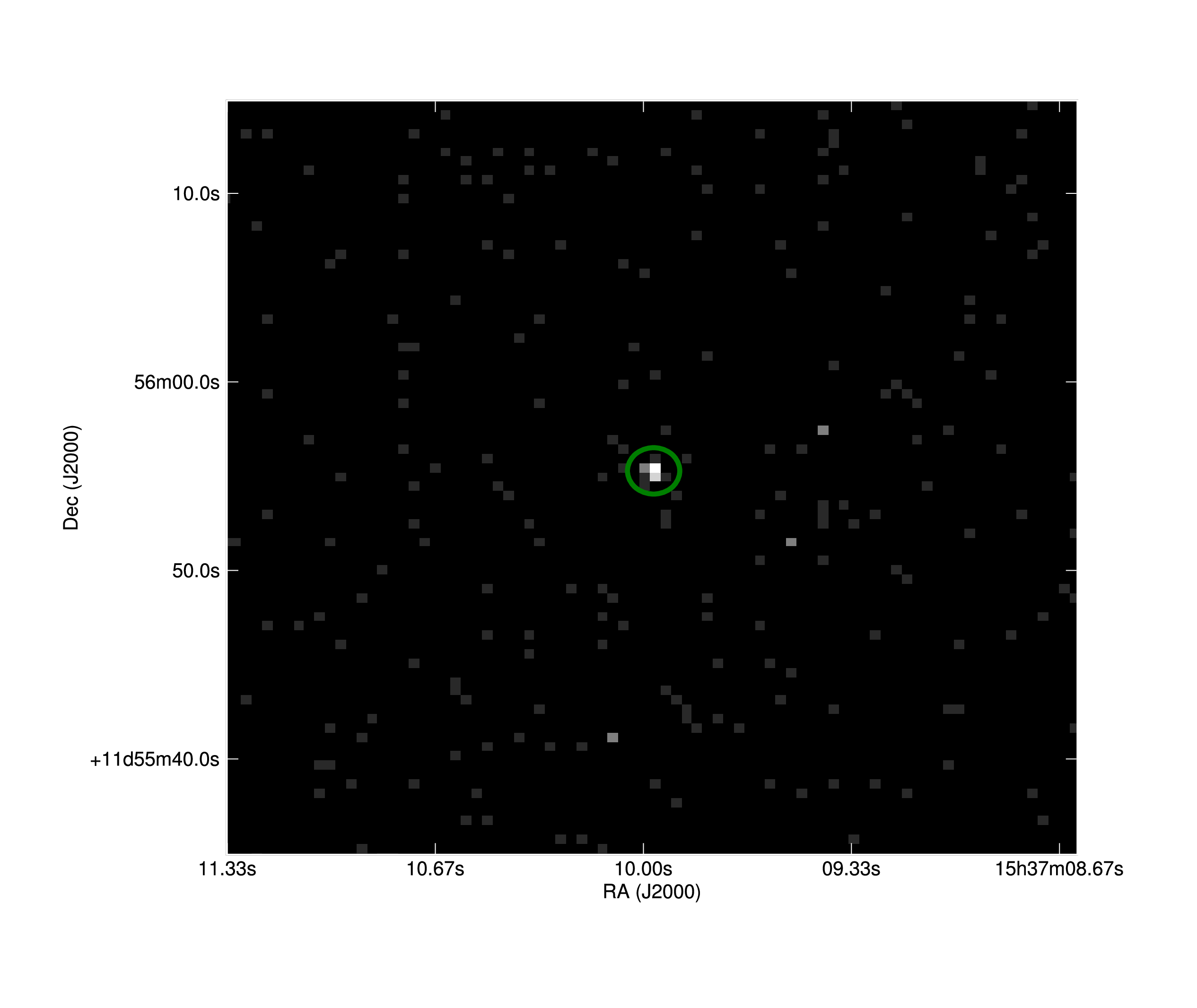}
\caption{40\arcsec$\times$40\arcsec\  image of the vicinity of J1537 in the 0.3--10\,keV range, obtained in the second  observation. The location of the $r=1\farcs23$ extraction circle is shown in green.
}\label{img}
\end{center}
\end{figure}

The second observation (ObsID 20224) was carried out on 2009 September 27 and had a total exposure time of 38181\,s. There were no bad time intervals. Again, the ACIS S3 chip was used,  this time in 128 row sub-array mode (i.e., 1/8 of the chip), 
 providing 441.04\,ms time resolution. The dead-time for such short 
 frame time is 9.3\%, so the total live  time after the dead-time correction is only 34628\,s. The less efficient high time resolution mode was used in order to make it possible to detect pulsations in the data, should slow radio pulsations from neutron star B be detected in the radio. As above, we extracted photons from an $r=1\farcs23$ 
  aperture  centered on the source. The total number of photons is 16, in the  
   0.3--10\,keV energy range. Of these,  on average, 0.69 photons are expected to come from the background. The image 
    from the second observation is shown in Figure \ref{img}. There appears to be no detector artifacts in the vicinity of the source, and the background beyond the extraction circle is quite uniform. 

\subsection{Timing}
\begin{figure}
\begin{center}
\includegraphics[width=\hsize]{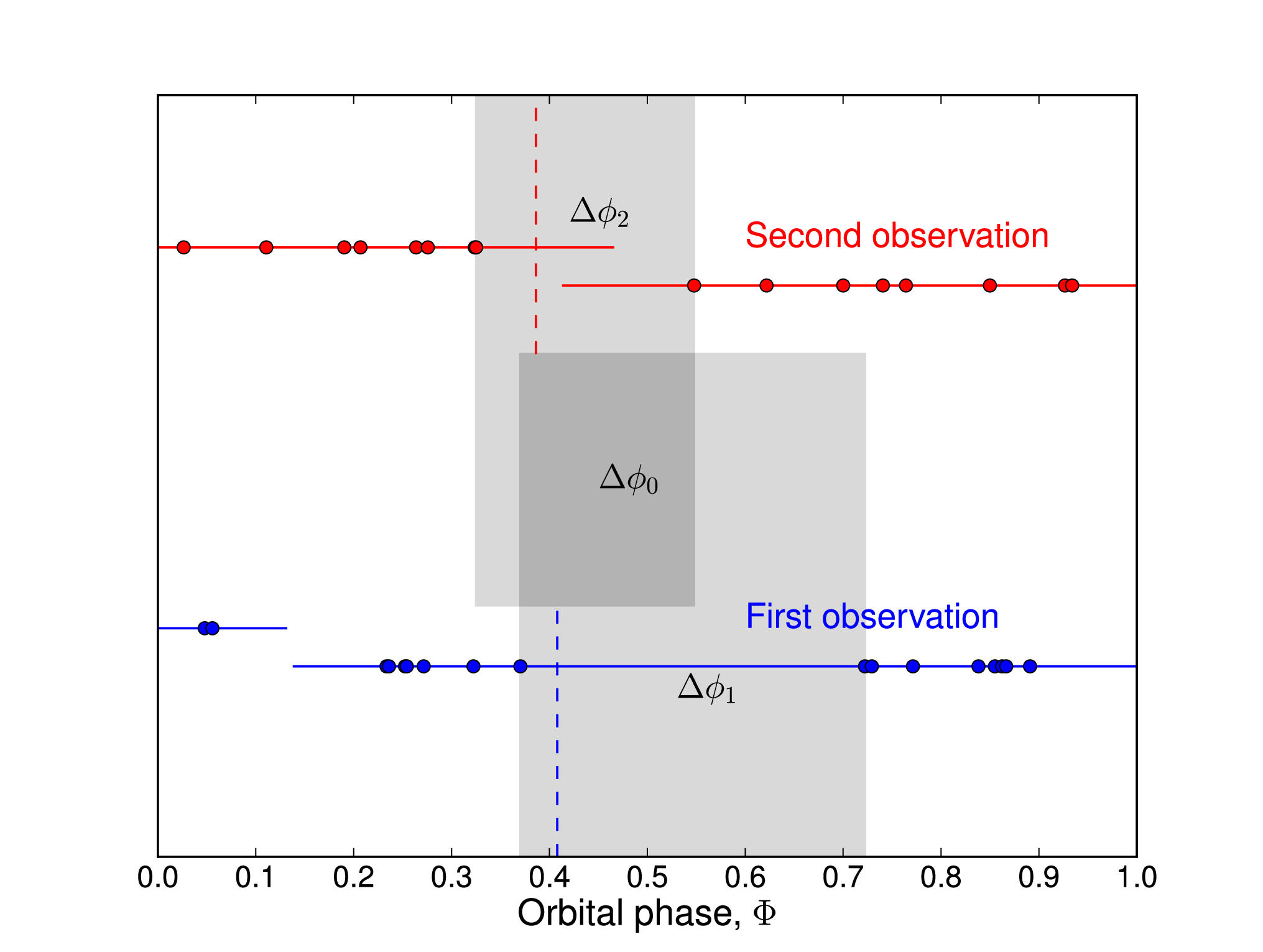}
\caption{Orbital phases of photon arrival times.
The horizontal lines show the phase intervals of the two observations.
The shaded areas show the phase intervals with no photons detected.
The vertical dashed lines show the phases of superior conjunction of pulsar A
(slightly different in the two observations because of the relativistic
advance of periastron).
}\label{phase}
\end{center}
\end{figure}

For the timing analysis, we used the photon arrival times from the Level 2 event files, after correcting them to the solar system barycenter with the {\tt axbary} tool from CIAO 4.2. For each photon, we calculated the orbital phase using the radio ephemeris (courtesy of Ingrid Stairs). The phase of each detected photon is plotted in Figure \ref{phase}. There appears to be a portion of orbital phase in which few/no photons are detected from the source.  We have checked that the fluctuations in the background rate were consistent with Poissonian statistics, and no decrease in the background rate was seen in the phase intervals with no counts detected in the source aperture. 

\begin{figure}
\begin{center}
\includegraphics[width=\hsize]{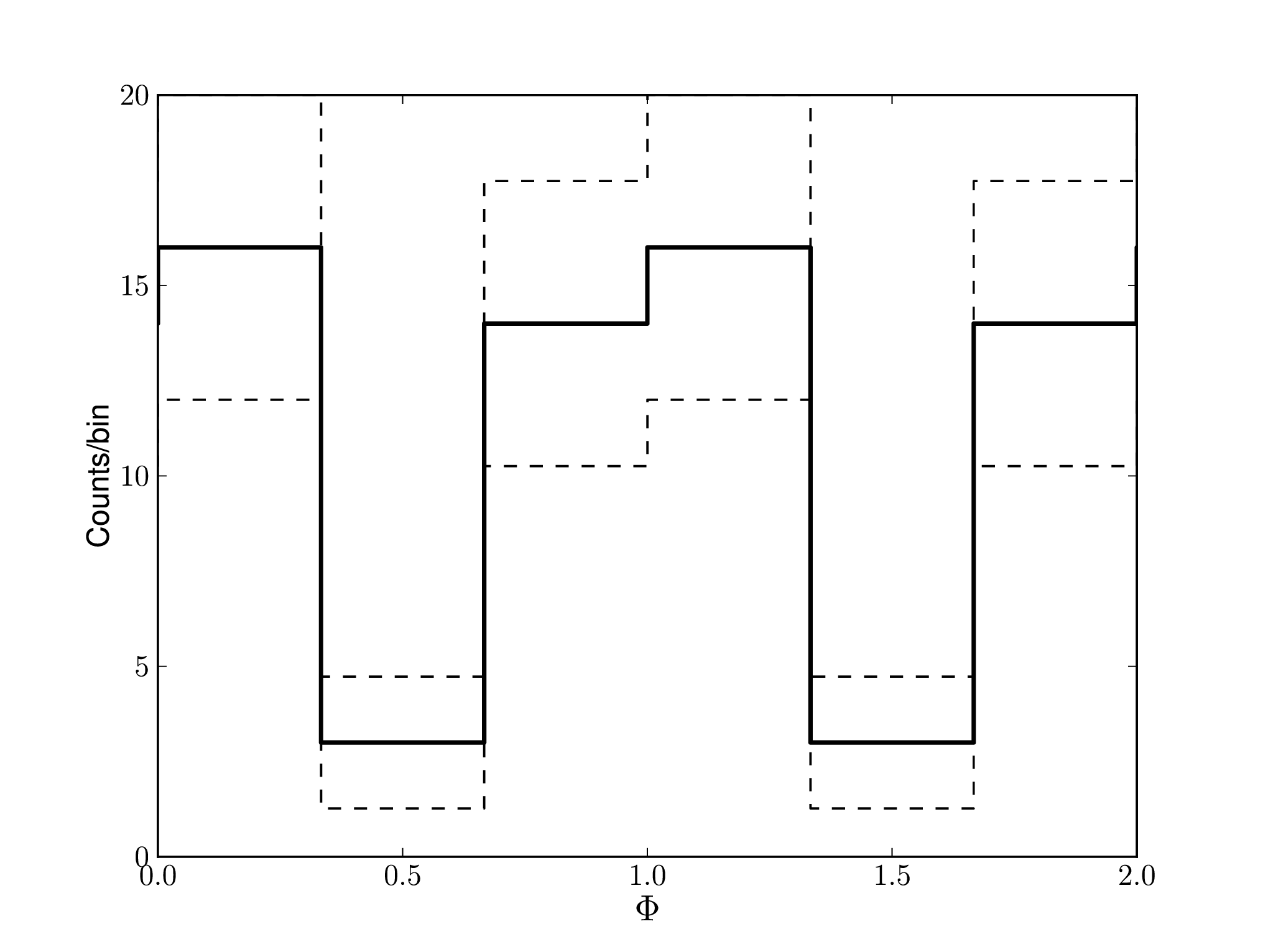}
\caption{Combined light curves for J1537 with three bins. The solid line is the actual counts histogram, whereas the dashed lines show the 68\% confidence interval for each bin, assuming Poissonian statistics. Two periods are shown for clarity.
}\label{lc}
\end{center}
\end{figure}

As an alternative representation of the same data, Figure \ref{lc} shows the phase-folded light curve, with three bins, where we choose one of the bins to be centered at the phase of apastron, $\phi_A=0.5$. We also show confidence intervals for each bin.

We will now  assess the probability that the observed deficit of photons  is caused by random fluctuations, assuming   that the number of counts within a certain fixed time interval obeys Poissonian statistics.

\begin{figure}
\begin{center}
\includegraphics[width=\hsize]{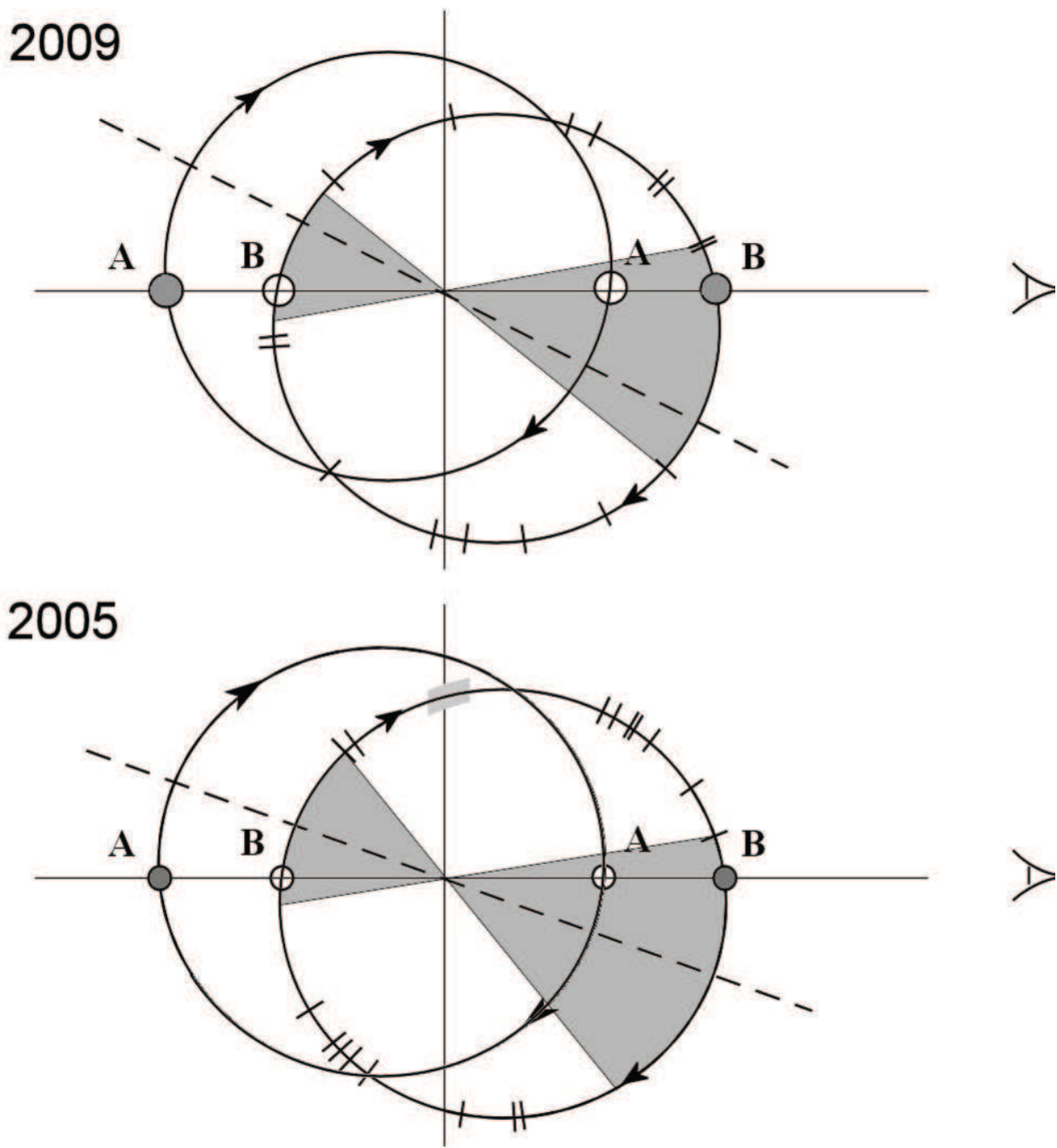}
\caption{Locations in the orbit where the detected photons were emitted: the tickmarks on the B orbital circle show the phases of the detected photons. The orbital configuration is shown for the two observation epochs, 2009 (top) and 2005 (bottom; as in KPG06). The solid vertical lines are the lines of nodes, and the dashed straight lines are the apsidal lines, whose orientations are different
in 2009 and 2005 because of
the relativistic precession. The locations of the conjunctions are shown with circles (filled for superior, open for inferior). The hatched region of the NS B orbit in the bottom panel shows the section of the orbit with no valid data. The shaded sectors show the ranges of true anomalies in which the lack of detected photons could be explained by NS B being out of the equatorial wind of pulsar A, because of a misalignment of the 
A's spin axis and the direction of the binary angular momentum
(see Section 3 and KPG06). The observer is to the right.
}\label{orbits}
\end{center}
\end{figure}

Assuming the null hypothesis that the photon arrivals are uniformly distributed in time, the probability of detecting $n$ photons  within a time interval $t$  is given by the Poissonian distribution $P_{\rm{Poiss}}(n,\lambda t)$, where $\lambda$ is the mean count rate (see Equation (A3) in the Appendix). In our case, it can be estimated as $\lambda=(N_1+N_2)/(T_1+T_2) 
= 16.154$ counts per binary period $P_{\rm orb}$, 
where $N_1=17$  and $N_2=16$ are the numbers of counts detected in the time intervals $T_1= 0.9925 P_{\rm orb}$ and $T_2=1.0503 P_{\rm orb}$, respectively, and $P_{\rm orb}= 36,352$ s.

The probability of finding an interval between two neighboring events larger than (or equal to) $\tau$ at a given event rate  $\lambda$ is given by Equations (A4) and (A1). For the first observation (in which the largest empty interval  is $\Delta t_6=0.352P_{\rm{orb}}$), these equations give the probability\footnote{Note that the probability $P_1$ was incorrectly estimated by KPG06 as $P_1=P_{\rm{Poiss}}(0,N_1 \Delta t_6/T_1)=0.34$\%.} $P_1(\tau_1\geq0.352P_{\rm{orb}})=3.81\%$. 

For the second observation, the largest intervals, $\Delta t_1=0.140 P_{\rm orb}$ and $\Delta t_{17} = 0.134 P_{\rm orb}$, are between the beginning of observation and the first count and between the last count and the end of observation. In this case, the exposure time in the gap is the sum of the two intervals at the start and end of the observations (see Figure \ref{phase}), $\Delta t_1 +\Delta t_{17}=0.274P_{\rm{orb}}$. The probability of obtaining such an interval by chance for uniformly distributed counts is estimated in the Appendix: $P_2(\tau_2\geq0.274P_{\rm{orb}})=19.0\%$.

We must assess the probability of finding this set of two gaps together, i.e.,  how likely is it that we will find by chance gaps as large or larger than observed  in two subsequent observations if the photons arrivals are uniformly distributed in time?  We have calculated the probabilities of  
 measuring a gap of a given (or larger) size within a single observation ($P_1$ and $P_2$) above.  An intuitive choice of  statistics for calculating the combined probability is then $Z=P_1P_2$ (e.g., \citealt{Fisher}). (Note that $P_1P_2$ is not itself the probability\footnote{For instance, for $n$ observations the corresponding product $\prod_i P_i$ will always tend to zero for large $n$ (even if the measured random process samples the null hypothesis) since each probability $P_i<1$.}).  Since $P_1$ and $P_2$ are both uniformly distributed, the cumulative distribution function is given by  $P(Z\leq z) = z -z \ln z $  (\citealt{SpringerBook}, page 102).
  In our case, $z=0.0072$ and  $P(Z\leq 0.007) =0.043$. Exactly the same probability follows from the recipe by  \citet{Fisher}: $P(Z<z) = P(\chi^2<\chi_0^2,\nu=2N)$, where the r.h.s.\ is the probability of observing a value of chi-squared larger than $\chi_0^2$ for $\nu$ degrees of freedom, $\chi_0^2 = -2 \ln z = 2 \ln\prod_{i=1}^N P_i$, and $N$ is the number of observations ($N=2$ in our case).


Finally, we account for the fact that the largest interval in the first observation and the largest interval   in the second observation  happen to overlap in  orbital phase space by a significant margin. For two  randomly chosen phase intervals $\Delta\phi'$ and $\Delta\phi''$, the probability of overlap by at least an amount $\phi_o$ is
\begin{equation}
	P_o=\Delta\phi'+\Delta\phi''-2\phi_o\,,
\end{equation}
for $0 < \phi_o < {\rm min}(\Delta\phi',\Delta\phi'')$. In our case,  $\Delta\phi'=0.352$, $\Delta\phi''=0.223$, and $\phi_o=0.177$ (see Figure 2), leading to the overlap probability $P_o=22.1$\%.
 
The combined probability of the two gaps being at least as large as observed and overlapping, is
\begin{equation}
P= P_{\rm comb}P_o = 0.22\times0.043 = 0.0094
\end{equation}
Thus, with a 99.1\% probability ($2.6\sigma$ significance), the observed two large, overlapping gaps suggest a physical dependence of the count rate on the orbital phase  rather than
a random fluctuation in the count rate of a steady source.
We confirmed the validity of the above calculations by extensive Monte-Carlo simulations.


Finally, we  used an alternative method based on the binned light curve (shown in Figure \ref{lc}) to estimate the probability of the observed count deficit. We produced the phase-folded light curve by binning  source counts from both observations.  Note that we chose the bin with the deficit to be centered  precisely on the physically significant phase of apastron, $\phi_A=0.5$, rather than picking a special phase which minimizes the number of counts  in the bin. The Poissonian probability of a bin containing only 3 or less photons when expecting 11 is 0.49\%. We can also evaluate the likelihood of such a light curve being a realization of a constant-rate model, with the expectation number of counts in each bin $e_i = 33/N$ for $N$ phase bins. Since the numbers of counts per bin  are  small,  the commonly used $\chi^2$ statistic (appropriate for large numbers of counts) is biased. Instead we can use the modified \citet{1979ApJ...228..939C} statistic
\begin{equation}
C = 2\sum_{i=1}^N n_i \times \left[ \frac{e_i - n_i}{n_i} - \ln\left(1 + \frac{e_i - n_i}{n_i}\right) \right]~,
\end{equation}
where $n_i$ is the measured number of counts in the $i$th bin. For 3 bins, we find $C=10.95$. To test how likely  it is to find a higher value of $C$  by chance, we generate 10000 sets of photon arrival phases, drawn from a uniform distribution in the 0--1 phase range. For each realization, we generate binned light curve as above, and calculate $C$. We find that $C>10.95$ in 0.4\% of cases. Thus, the null hypothesis of a count rate independent of phase is ruled out with 99.6\% confidence. The apparently higher significance reflects our choice of the number of bins in the light curve.

To visualize the placement of the observed deficit in relation to the NS orbits and the observer's line-of-sight,  we plotted in Figure 4
 the configurations of the orbits and marked the true anomaly values corresponding to the photon arrival phases.
The relativistic advance of periastron by $\delta\omega=\dot{\omega}T=1.756\degr$\,yr$^{-1}\times4.465$\,yr\,$=7.84\degr$ is accounted for. 

\subsection{Spectrum}
\begin{figure}
\begin{center}
\includegraphics[width=\hsize]{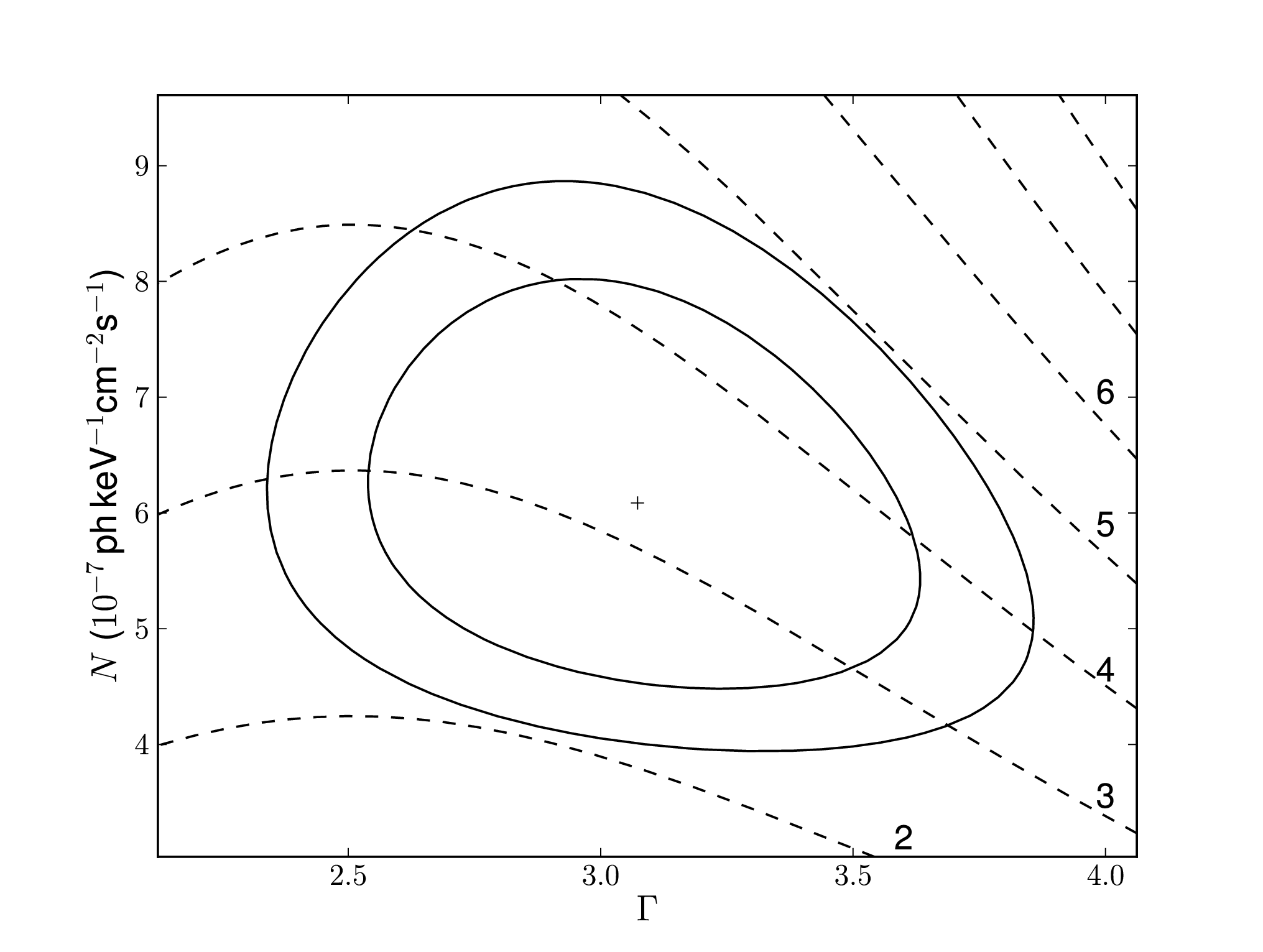}
\caption{Confidence contours (68\% and 90\%)
  for absorbed PL model fit to both data sets jointly, for photon index $\Gamma$ and normalisation $\cal{N}$. Dashed lines show loci of constant unabsorbed flux, labelled by flux values in units of $10^{-15}$\,erg\,s$^{-1}$cm$^{-2}$, for 0.3--8\,keV.
}\label{conf}
\end{center}
\end{figure}

We extracted the spectra and produced  response files following standard ACIS data reduction procedure. Spectral analysis was performed with  {\sl Sherpa}\footnote{See {\tt http://cxc.harvard.edu/sherpa/}}  in the 0.3--8\,keV range. Due to the low numbers of photons, we fitted the models by minimizing the C-statistic (Cash 1979) without binning. We fitted a power-law (PL) model, modified by interstellar absorption (equivalent to {\tt wabs*powerlaw} in XSPEC), and  fixed the  absorbing column to  $N_H=3.6\times10^{20}$\,cm$^{-2}$. This value corresponds to the pulsar's dispersion measure for 10\% ionization, and it is close to the H\,{\sc i} column density through the Galaxy in this direction. Note that such a low extinction makes virtually no effect on the spectrum above $0.3$\,keV. We find  that, between the two data sets, the best-fit parameters  are  consistent within their 1$\sigma$ uncertainties   (see Table \ref{fit}).   The combined fit     results are  shown in Table \ref{fit}, and the confidence contours corresponding to these parameters  are plotted in Figure \ref{conf}. The unabsorbed luminosity  is $L_{\rm 0.3-8\,keV}=(4.0\pm1.0)\times10^{29}$\,erg\,s$^{-1}$, for the distance of 1.02\,kpc.

We also jointly  fitted an absorbed blackbody (BB) model and obtained a slightly worse fit 
 ($C=154$ vs.\ $C=127$ for the PL). The best fit BB parameters,  
 $kT=262^{+30}_{-18}$\,eV and $R=20^{+15}_{-9}$\,m,
 are at the hotter/smaller end of the confidence regions calculated by KPG06.
  We cannot exclude the possibility that the spectrum is described by
a multi-component model, but the statistics are too poor to analyze such models. 

\begin{deluxetable}{cccc}
\tablecaption{Spectral fits\label{fit}}
\tablewidth{0in}
\tablehead{
\colhead{Parameter} & \colhead{Old\tablenotemark{a}} & \colhead{New}  & \colhead{Joint}
}
\startdata 
$\Gamma$ 		& 2.9$\pm$0.5		& 3.3$\pm0.6$		&3.07$^{+0.37}_{-0.36}$\\
${\cal N}$ \tablenotemark{b}	& 6.4$^{+1.8}_{-1.5}$	& 5.7$^{+1.8}_{-1.5}$	
& $6.1^{+1.2}_{-1.1}$ \\
$N_H$ ($10^{20}$\,cm$^{-2})$ & 3.6		& 3.6			&3.6\\
$F_{0.3-8\,\small\textrm{keV}}$ \tablenotemark{c}& 3.2$\pm$0.9& 3.3$\pm$1.1	&3.2$\pm$0.8\\
$C_{\rm red}$\tablenotemark{d}		& 0.14			& 0.10			&~~0.12
\enddata
\tablecomments{All fits are performed in the 0.3--8\,keV energy range, 
the background ignored, and C-statistics are used without binning. $N_H$ is held fixed.}
\tablenotetext{a}{Fit slightly different from KPG06, due to  the newer callibration used and an extra photon included.}
\tablenotetext{b}{Normalization in units of $10^{-7}\,$photons\,keV$^{-1}$cm$^{-2}$s$^{-1}$ at 1\,keV.}
\tablenotetext{c}{Flux in units of $10^{-15} $\,erg\,s$^{-1}$cm$^{-2}$, in 0.3--8\,keV range.}
\tablenotetext{d}{C-statistic \citep{1979ApJ...228..939C} per detector energy channel.  }
\end{deluxetable}

\section{Discussion and Conclusions}
 KPG06 suggested three possible scenarios to explain the  X-ray emission from J1537: (a) the emission is generated in the magnetosphere or heated polar caps of the recycled  pulsar A; (b)  the emission results from interaction of the pulsar A's  wind with the NS companion, such that the strength of the interaction  depends on the varying  distance between the two NSs in the eccentric orbits; and (c)  emission only happens when the companion NS B passes through the equatorial outflow of the A's wind. Model (a) predicts no dependence on orbital phase, (b) predicts fainter emission at apastron and brighter emission at periastron, and (c) predicts two segments with faint/no emission at opposite phase intervals, around apastron
and periastron. Note that modulation would be possible for scenario a) via absorption in the magnetosphere of NS B; however with the small magnetosphere of NS B and the inclination of the orbit, it is impossible for this system. We find a deficit of photons around apastron, and no evidence for enhanced emission at periastron. Thus,
 our data appear to support model (c).
As the pulsar's equatorial plane 
and the orbital plane are misaligned 
(by $\delta =25^\textrm{o}\pm4^\textrm{o}$; \citealt{2004PhRvL..93n1101S}), the 
NS B will intersect the wind for only part of its orbit, consisting
of two segments (see Figure 4). One expects more emission as the 
NS B plunges through the pulsar wind, and less when it 
 is outside the wind. Following the calculation in KPG06, the current results 
support a limit on the equatorial outflow opening angle of about $\Delta\geq20^\textrm{o}$. X-rays can be generated in the intra-binary  shock or can originate from the J1537B's
magnetosphere or surface, bombarded by the particles of J1537A's wind.

We should note that the phases of superior conjunction of pulsar A (slightly different
in the two observations because of the relativistic advance of periastron)
are  also within the
$\Delta\phi\approx 0.18$ phase interval in which no photons were detected
in the two observations (see Figures 2 and 4). 
However, since
an excess of flux around superior conjunction
 would be more natural than a deficit,
due to the Doppler boost,
we believe that the observed deficit is connected with NS B being outside
of the pulsar A's wind around apastron, and is not related to superior
conjunction.

In the light of this interpretation, it is illustrative to compare the X-ray properties of J1537 and J0737, the only other DNSB detected in X-rays. For J0737, the 
pulsed flux from the recycled pulsar J0737A
   accounts for 
$\approx 76\%$ 
of the total DNSB flux in the energy range 
0.15--4\,keV, 
the pulsed fraction growing with energy \citep{2008ApJ...679..664P}. There is also some evidence for pulsations from the slower, non-recycled pulsar J0737B, 
  but only for  small ($\Delta\phi=0.25$) part of the orbit,
  and the pulsed fraction is rather uncertain.
Measuring pulsations with the J1537A's period in the J1537 X-ray emission
would provide a lower limit on the J1537A's contribution to the total observed
X-ray flux, but the low time resolution of the {\sl Chandra} ACIS observations
does not allow such a measurement.

In contrast with J1537,
no orbital phase dependence is seen in the X-ray emission of J0737, 
down to the limit of $\lesssim15$\% on pulsed fraction \citep{2008ApJ...679..664P}. 
It itself does not exclude a sizeable fraction of the J0737 luminosity to
be powered by the J0737A's wind. Indeed,
both the eccentricity ($e=0.0877$; \citealt{2008ApJ...680..654P})
and the angle between the orbital angular momentum and the A's spin 
 ($<14^{\circ}$ at 95\% confidence; \citealt{2008AIPC..983..474F}) 
are much smaller than those in J1537.
As a result, the distance between the pulsars changes very little,
  and pulsar J0737B  
 is  always near the equatorial plane of J0737A, i.e., within the wind of  J0737A, even if the opening angle of the equatorial  outflow is small. One would not, therefore, expect the emission caused by the interaction
of the J0737A's wind with J0737B
 to vary significantly with orbital phase
(except, possibly, for variations caused by the changing view of the shock 
front). 

The spectra of both J1537 and J0737 are rather soft in comparison with  a typical pulsar wind nebula \citep{2008AIPC..983..171K}, and their slopes,
$\Gamma\approx3.1$ and  $\Gamma\approx3.4$ (for the phase-averaged spectrum\footnote{\citet{2008ApJ...679..664P} also performed  phase-resolved spectroscopy of J0737A and concluded that three components,  two  BBs (with $kT\sim150$\,eV and $kT=30-85$\,eV) and a PL  (with $\Gamma=3.3^{+0.1}_{-0.2}$), are required to simultaneously fit the spectra in three phase intervals. The modulation of the PL normalization was by itself able to explain the large pulsed fraction ($75.7\%\pm5.4\%$ in the 0.15--4 keV band)
 at J0737A's period.}; \citealt{2008ApJ...679..664P}), respectively, 
are at the upper end of the $\Gamma\simeq2$--3 range typical for 
recycled and non-recycled old pulsars 
whose spectra can often be approximated by an absorbed PL if  the number of counts is small\footnote{The spectra of ordinary old
and recycled pulsars with 
high S/N usually require three components for an acceptable fit, two thermal components (BB or NS atmosphere) and a PL. }  (e.g., \citealt{2010ApJ...709..241B};  \citealt{2009ApJ...691..458P}, and references therein). 
The similarity of the J1537 and J0737 spectra
seems contradicting the 
interpretation of the J1537 X-ray emission as powered by the J1537A's wind 
because the X-ray luminosities of the two DNSBs are dominated by physically 
different mechanisms.
However, even though most spectra of PWNe around isolated pulsars 
  fit harder PLs  \citep{2008AIPC..983..171K}, we  know very little about emission from pulsar winds  shocked in the vicinity of their parent pulsar due to the presence
   of a binary companion. It is possible that spectra of intrabinary PWNe are softer than those of ordinary PWN (e.g., due to a smaller
 amount of the magnetic field energy
 being transferred to the particles, which also implies a higher magnetization, $\sigma$; cf.\ \citealt{2008ApJ...679..664P}). 
We also note that the J1537's spectrum is marginally harder
 than that of J0737,
 and the large uncertainties leave room for even more appreciable differences. 
Deeper X-ray observations of J1537 should better constrain its spectrum and orbital dependence,

A significant difference 
between the two DNSBs
is that the 
X-ray radiative efficiency, 
$\eta_{\rm 0.3-8\,keV} = L_{0.3-8\,\small\textrm{keV}}/\dot{E}_A \approx2.2\times10^{-4}$
for J1537, is a factor of 
7 lower than for J0737 (with account for the revised distance to J0737,
$d\simeq 1150$\,pc; \citealt{2009Sci...323.1327D}).
  The X-ray efficiency of J0737, $\eta_{\rm 0.3-8\,keV}\approx 1.6\times 10^{-3}$, is 
typical for millisecond pulsars detected in X-rays
(see Figure 4  in \citealt{2007ApJ...664.1072P}),
in accordance with the conclusion that
the  J0737's emission is dominated by pulsar J0737A \citep{2008ApJ...679..664P}.
   On the contrary, the J1537's efficiency is lower than those of 
   nearly all other recycled pulsars.
If most of the X-ray luminosity in J1537 comes from the interaction of
the  pulsar A's wind with pulsar B,
as follows from our interpretation,
the efficiency of  J1537A is even lower,
making it one of the least efficient recycled pulsars known.

An upper limit on the wind-powered
J0737's luminosity,
$\sim 1.7\times 10^{30}$ erg s$^{-1}$ in the 0.3--8 keV band
(which includes the J0737B's luminosity), can be
estimated as the difference between the total observed luminosity and the
luminosity pulsed with the A's period. 
This limit is a factor of 4 larger than the total luminosity of J1537, 
but this fact does not exclude the possibility that 
the unpulsed
luminosity of J0737 is wind-powered because
the J0737 orbit is tighter. 

 In summary, we have confirmed the 
  deficit of  X-ray emission in the J1537 DNSB around 
   apastron. This implies 
that a significant fraction of the X-ray emission is powered by the
equatorial wind of the recycled pulsar J1537A, 
and the emission probably occurs when
 the companion NS crosses this wind.
    Our interpretation also implies that the apparent X-ray efficiency of pulsar J1537A is very low compared to other millisecond pulsars.

\acknowledgements
We are deeply indebted to Ingrid Stairs for providing us with the up-to-date radio ephemeris. 
 This 
 work was supported by {\sl Chandra} award GO9-0073X and NASA grant
 NNX09AC84G.
The research was also partly supported by the National Science Foundation under Grants No.\ 0908733 and 0908611. The work by GGP was partly supported by the
Ministry of Education and Science of Russian Federation (Contract No.\ 11.G34.31.0001).

{\it Facilities:} \facility{Chandra (ACIS)}, \facility{XMM-Newton (EPIC)}

\bibliography{database}

\appendix

\section{Probability calculation}

The probability 
that, for an observation
with $N$ counts detected,
the largest of the $N+1$ time intervals between neighboring events 
(which include the beginning and the end of the observation)
is  as large as or larger than $\tau$
 can be calculated as 
the difference between unity and the probability of all the intervals being
smaller than $\tau$:
\begin{eqnarray} \label{eq2}
P(\Delta t\geq \tau\,|\,N;T)= \nonumber\\
1-N!T^{-N}
\int_0^\tau d(\Delta t_1)\ldots \int_0^\tau d(\Delta t_N)\,\,
\Theta\left(T-\sum_{i=1}^{N}\Delta t_i\right)\,\Theta\left(\sum_{i=1}^{N}\Delta t_i+\tau -T\right),
\end{eqnarray}
where $T$ is the duration of the observation,  $\Theta(x)$ is the Heaviside step-function ($\Theta(x>0)=1, \Theta(x<0)=0$), 
and $\Delta t_i$ is the interval between the $(i-1)$th and $i$th events ($i=0$ and $i=N+1$ correspond to the beginning and the end of the observation, respectively).
Equation (A1) takes into account that, for $N$ fixed,
each possible configuration of intervals ought to have the same probability of occurrence. The step-functions in Equation (A1) correspond to the following constraint
\begin{equation}
0 < T -\sum_{i=1}^N \Delta t_i = \Delta t_{N+1} < \tau\,.
\end{equation}
The analytical calculation of the $N$-dimensional integral in (A1) leads to very cumbersome expressions; in particular, for each given $N$, they are different for the separate intervals $(k+1)^{-1} < \tau/T < k^{-1}$, where $k=1,2,\ldots N$. However, the integral can be evaluated numerically.

The number of counts is itself a random quantity that obeys some statistical distribution. For the null hypothesis we are investigating, the probability of detecting $n$ photons in a time interval $t$ is given by the Poissonian distribution
\begin{equation}
P_{\rm Poiss}(n,\lambda t)=\frac{(\lambda t)^n\, e^{-\lambda t}}{n!}\,,
\end{equation}
where $\lambda$ is the mean count rate.
For a given $\lambda$, the probability that the largest interval between 
neighboring events is as large as or larger than $\tau$ is the
expectation value of $P(\Delta t\geq\tau\,|\,N;T)$:
\begin{equation}
P(\Delta t\geq\tau\, |\, \lambda T)=\sum_{N=0}^{\infty} P(\Delta t\geq\tau\, |\, N;T)\, P_{\rm{Poiss}}(N,\lambda T)\,
\end{equation}
For the first observation ($\tau=0.352$, $T=0.9925$, $\lambda = 16.15$; the unit of time is the orbital period) this gives $P(\Delta t > 0.352\,|\,16.03)\equiv P_1 = 0.0381$.

For the second observation, the largest interval in the phase space is not the one between the consecutive events  but  the sum of the waiting times before the first count and after the last one (see Figure 2). 
The probability of the sum of two intervals to be larger than $\tau$ when the total number of events $N$ in the observation time $T$ is given can be calculated as  
   \begin{eqnarray} \label{eq_p2}
P(t_1+t_{N+1}\geq\tau\, |\, N;T)=\nonumber\\
1 - N! T^{-N}
\int_0^\tau d(\Delta t_1)\ldots \int_0^\tau\, d(\Delta t_N)\,\,
\Theta\left(T-\sum_{i=1}^{N}\Delta t_i\right)\,\,\Theta\left(T-\sum_{i=2}^{N}t_i-\tau\right),
\end{eqnarray} 
 Since the expectation value of  $P(t_1+t_{N+1}\geq\tau\, |\, N;T)$ is  
$P(t_1+t_{N+1}\geq\tau\, |\, \lambda T) = \sum_{N=0}^{\infty} P(t_1+t_{N+1}\geq\tau\, |\, N;T)P_{\rm{Poiss}}(N,\lambda T)$, for the second observation we obtain $P(t_1+t_{17} \geq 0.274\, |\, 16.70) \equiv P_{2}=19.0$\%.

\end{document}